\documentstyle[times]{mn}
\input{epsf}
\newcommand{\bez}{\begin{eqnarray*}}
\newcommand{\eez}{\end{eqnarray*}}
\newcommand{\be}{\begin{equation}}
\newcommand{\ee}{\end{equation}}
\newcommand{\beq}{\begin{eqnarray}}
\newcommand{\eeq}{\end{eqnarray}}
\newcommand{\bc}{\begin{center}}
\newcommand{\ec}{\end{center}}

\newbox\grsign \setbox\grsign=\hbox{$>$} \newdimen\grdimen \grdimen=\ht\grsign
\newbox\simlessbox \newbox\simgreatbox \newbox\simpropbox
\setbox\simgreatbox=\hbox{\raise.5ex\hbox{$>$}\llap
     {\lower.5ex\hbox{$\sim$}}}\ht1=\grdimen\dp1=0pt
\setbox\simlessbox=\hbox{\raise.5ex\hbox{$<$}\llap
     {\lower.5ex\hbox{$\sim$}}}\ht2=\grdimen\dp2=0pt
\setbox\simpropbox=\hbox{\raise.5ex\hbox{$\propto$}\llap
     {\lower.5ex\hbox{$\sim$}}}\ht2=\grdimen\dp2=0pt
\def\simgt{\mathrel{\copy\simgreatbox}}
\def\simlt{\mathrel{\copy\simlessbox}}

\def\ep{\varepsilon}
\def\gga{\gamma-\gamma}
\def\bbeta{\mbox{\boldmath $\beta$}}
\def\bv{\mbox{\boldmath $v$}}

\begin{document}

\title[Electron-positron outflows]
{Electron-positron outflows from gamma-ray emitting accretion discs}

\author[A.~M. Beloborodov]
{\parbox[]{6.8in} {A.~M. Beloborodov$^\star$}\\
Stockholm Observatory, S-133 36 Saltsj\"obaden, Sweden}

\date{Accepted, Received}

\maketitle

%This is single spacing
%\baselineskip 12pt
%This is double spacing
%\baselineskip 24pt

\begin{abstract}
An $e^\pm$ atmosphere is inevitably created around a black hole 
accretion disc, the spectrum of which extends to MeV energies. 
Pairs created in $\gamma-\gamma$ collisions outside the disc 
are blown away by soft radiation 
(which dominates the bolometric luminosity of the disc) and
form a semi-relativistic outflow. We simulate numerically the conversion of 
the MeV radiation into a vertical $e^\pm$ outflow above a disc-like source.
The outflowing $e^\pm$ plasma  becomes optically thick to Thomson scattering 
if the compactness of the $\gamma$-ray source exceeds $\sim 30$.
The scattering by $e^\pm$ then collimates the bulk of  
soft radiation along the disc axis, and the apparent bolometric 
luminosity of the disc depends strongly on its inclination to the line of sight.
The anisotropic central emission may account for the lack of Fe K$\alpha$ 
lines in the X-ray spectra of bright radio-quiet quasars. 
The scattering in $e^\pm$ outflows may also explain the orientation of 
optical polarization in non-blazar active galactic nuclei. 
\end{abstract}


\section{Introduction}

The X-ray activity of  
galactic black hole candidates (GBHs) and active galactic nuclei (AGNs) 
is usually attributed to accretion discs  surrounding putative black 
holes. A special feature of accreting black holes is that their radiation 
spectra extend up to the $\gamma$-ray band. The $\gamma$-rays are thought to be 
produced
in the innermost region of the accretion disc (see, e.g., Svensson 1996 for a 
review). It may be a cold disc covered by an active corona in which particles 
are accelerated in magnetic reconnection events (Galeev, Rosner \& Vaiana 
1979). It may also be a hot two-temperature disc as proposed by Shapiro, 
Lightman \& Eardley (1976).

\footnotetext{$^\star$Also at Astro Space Centre of Lebedev Physical 
Institute, 84/32 Profsojuznaja Street, Moscow 117810, Russia}

As soon as the radiation spectrum extends above the electron rest-mass energy,
$m_ec^2=511$ keV, the reaction 
$\gamma+\gamma\rightarrow e^-+e^+$ becomes inevitable and leads to production 
of an electron-positron atmosphere of the black hole (Herterich 1974). 
Guilbert, Fabian \& Rees (1983) showed that if
the $\gamma$-ray luminosity exceeds $\sim 10^{-3}$ of the Eddington limit for
the black hole, the $e^\pm$ atmosphere 
becomes optically thick to Thomson scattering. 
The brightest accretion discs thus can be surrounded by optically thick 
$e^\pm$ envelopes. 
% This is likely to occur in quasars whose luminosity is 
% comparable to the Eddington limit. 
The dynamics of an $e^\pm$ envelope around a compact $\gamma$-ray source 
has been discussed in various contexts (e.g., Katz 1982; Zdziarski 1984; 
Guilbert \& Stepney 1985; Coppi \& Lamb 1992; Illarionov \& Krolik 1996; 
Li \& Liang 1996),
and it has been shown that the pairs should share the momentum of radiation 
and form an outflow from the source. 

In this paper, we study the formation of vertical $e^\pm$ outflows from   
disc-like $\gamma$-ray sources. Following the 
observed spectra of non-blazar AGNs and GBHs, we assume that the $\gamma$-rays 
contribute only a small fraction, $\simlt 0.1$, to the 
bolometric luminosity, and the bulk of disc emission is in
UV or X-ray bands. The created pairs then move in the dense soft radiation
field which forces them to acquire an equilibrium velocity $\sim c/2$ such that 
the effective radiation pressure acting on the pairs vanishes 
(Gurevich \& Rumyantsev 1965). The equilibrium
velocity is determined by the angular distribution of the radiation and can
easily be calculated for an optically thin atmosphere.
In the optically thick case, the problem becomes non-linear as  
scattering by moving pairs affects the angular distribution 
of the radiation above the disc. 
The corresponding transfer problem can be solved self-consistently to yield 
both the plasma velocity and radiation intensity as a function of optical depth
in the $e^\pm$ atmosphere (see Section 4). 

The density of the outflow is governed by pair injection due to $\gamma-\gamma$
interactions. We assume that the primary spectrum of $\gamma$-rays emerging from
the source has a break at a few MeV due to strong absorption
of more energetic photons inside the source. The bulk of $\gga$ interactions
outside the disc occur between MeV photons, just above the threshold.
The $\gamma$-ray transfer is therefore non-linear as the $\gga$ 
opacity at each height is determined by the MeV radiation field itself.
To study the non-linear transformation of $\gamma$-rays into an $e^\pm$ 
outflow, we have developed a code which calculates the hard X/$\gamma$-ray 
transfer in a slab geometry. Besides the $\gga$ reaction, the code 
approximately accounts for Compton scattering and annihilation 
emission of the created pairs, assuming that the pairs move with the typical 
velocity $v=c/2$. 

The main properties of the atmosphere, 
such as Thomson optical depth and the power of $e^\pm$ outflow,
are determined basically by one parameter of the source -- the 
compactness parameter defined in Section 2.1.
The results are weakly dependent on the specific internal boundary conditions 
for the transfer problem.
In our calculations, the primary source is simulated in the 
simplest way: A homogeneous disc emitting hard photons
isotropically with a power-law spectrum around 511 keV.

In Section 2, we discuss a general constraint imposed by $\gga$ absorption 
on the spectrum of an accretion disc. Then, in Section 3, we calculate the
rate of pair injection above the disk surface and discuss the formation of 
an optically thin $e^\pm$ outflow. In Section 4, we consider
$\gamma$-ray sources with large compactness parameters, which 
are surrounded by optically thick $e^\pm$ atmospheres, and calculate
the $e^\pm$ density and velocity profiles of the optically thick outflows.
We also discuss pair acceleration at large distances from the source.
The main conclusions are summarized in Section 5.

%#############################################################################

\section{Gamma-gamma absorption}

$\gga$ absorption imposes a restriction on the maximum energy
of the $\gamma$-rays emitted by a compact source (Guilbert et al. 1983). 
This provides a test for whether the observed $\gamma$-rays can be emitted by 
an accretion disc or if they should be associated 
with a relativistic jet or a more complicated geometry of the source 
(Illarionov \& Krolik 1996). Such a test has been applied to $\gamma$-loud AGNs 
(e.g., McNaron-Brown et al. 1995; Becker \& Kafatos 1995).

\subsection{The model}

Consider a homogeneous $\gamma$-ray emitting disc of radius $R$ 
and assume that the hard X/$\gamma$-ray emission emerging from the 
disc is isotropic and has a power-law spectrum around 511 keV 
with intensity (flux per unit energy per steradian) 
\begin{equation}
   I_\ep=I_1\ep^{-\alpha},
\end{equation}
where $\ep=h\nu/m_{\rm e}c^2$ is photon energy in units of the electron rest 
mass energy. The disc spectral luminosity is given by
\begin{eqnarray}
\nonumber
   L_\ep=L_1\ep^{-\alpha}, \qquad L_1=2\pi^2R^2I_1.
\end{eqnarray}
The main parameter of the model is the compactness parameter,
\begin{equation}
    l_1=\frac{L_1\sigma_{\rm T}}{m_{\rm e}c^3R},
\end{equation}
which can also be expressed in terms of the Eddington luminosity, 
$L_{\rm E}=2\pi r_{\rm g} m_{\rm p} c^3/\sigma_{\rm T}$, 
\begin{eqnarray}
\nonumber
   l_1=2\pi\frac{m_{\rm p}}{m_{\rm e}}\frac{L_1}{L_{\rm E}}\frac{r_{\rm g}}{R},
\end{eqnarray}
where $r_{\rm g}=2GM/c^2$ is the gravitational radius of a black hole of mass
$M$. 
We assume the typical spectral index $\alpha>1$, then the $\gamma$-ray 
intensity steeply falls off towards high energies and $l_1$ roughly 
corresponds to the total luminosity above 511 keV. This luminosity is 
a few per cent of the bolometric luminosity of the disc in UV and X-rays.

\subsection{Absorption inside and outside the source}

For a rough estimate of $\gga$ absorption inside the source, take the $\gga$ 
opacity of an isotropic radiation field (Gould \& Schr\'{e}der 1967) 
\begin{equation}
 \kappa_{\gamma\gamma}^{\rm isotrop}(\ep)=
  \eta\sigma_{\rm T}\frac{w(\ep^{-1})}{m_{\rm e}c^2}=
   \frac{2\eta l_1}{\pi R}\;\ep^\alpha,
\end{equation}
where $w(\ep)=4\pi I_\ep/c$ is the spectral energy density of the
radiation, and $\eta(\alpha)$ is a numerical factor, e.g.,
$\eta\approx$ 0.122, 0.072, 0.047 for $\alpha$=1, 1.5 , 2 respectively. 
The position of the spectral break due to absorption inside the source,
$\ep_{\rm max}$, can be estimated from the condition
$\kappa^{\rm isotrop}_{\gamma\gamma} h \sim 1$, where $h$ is the 
disc thickness. Then one gets 
\begin{equation}
   \ep_{\rm max}\sim \left(\frac{\pi}{2\eta l_1}\frac{R}{h}\right)^{1/\alpha}.
\end{equation}
In the case $h\ll R$ this is a much weaker constraint than
that for a spherical source with $h\sim R$. 
However, equation (4) is not the final answer to the problem of $\gga$
absorption because those $\gamma$-rays which have not been absorbed inside 
the disc-like source may be absorbed above the disc when passing through its
radiation field (see also Zdziarski 1984).

Let us evaluate absorption for a $\gamma$-photon of energy $\ep$ emitted along 
the disc 
axis ($z$-axis) assuming for simplicity that the disc compactness is small 
($l_1<1$) and the non-linear effects are unimportant. Then at height $z$ the 
$\gga$ opacity seen by our photon is caused by the radiation propagating freely 
from the disc within an angle $\theta_{\rm max}=\arctan(R/z)$.
Photons of energy $\ep^\prime$ streaming 
at angle $\theta<\theta_{\rm max}$ can interact with our
photon if $\ep^\prime$ exceeds the threshold
\begin{equation}
  \ep_{\rm thr}=\frac{2}{\ep(1-\cos\theta)}.
\end{equation}
The $\gga$ opacity is then given by
\begin{eqnarray}
\nonumber
  \kappa_{\gamma\gamma}(\ep,z)=\frac{l_1}{\pi\sigma_{\rm T} R}
  \int_0^{\theta_{\rm max}} {\rm d}\theta\sin\theta\int_{\ep_{\rm thr}
                                         (\theta,\ep)}^
  {\ep_{\rm max}} {\rm d}\ep^\prime {\ep^\prime}^{-\alpha-1} \\
  \times (1-\cos\theta)\; \sigma_{\gamma\gamma}(\ep,\ep^\prime,\theta),
\end{eqnarray}
where $\sigma_{\gamma\gamma}$ is the cross section for $\gamma-\gamma$
pair production (Jauch \& Rohrlich 1976)
\medskip
\begin{eqnarray}
\nonumber
\sigma_{\gamma\gamma}(\ep_{\rm c})=\frac{3\sigma_{\rm T}}{8\ep_{\rm c}^2}
 \left[
 \left(2+\frac{2}{\ep_{\rm c}^2}-\frac{1}{\ep_{\rm c}^4}\right)
 \ln\left(\ep_{\rm c}+\sqrt{\ep_{\rm c}^2-1}\right) \right.  \\
 \left.-\left(1+\frac{1} {\ep_{\rm c}^2}\right) 
  \sqrt{1-\frac{1}{\ep_{\rm c}^2}}\;
 \right],
\end{eqnarray}
$\ep_{\rm c}=(\ep^\prime/\ep_{\rm thr})^{1/2}$ is the energy of the 
interacting photons in the center-of-momentum frame.
After integration we find
\begin{equation}
 \kappa_{\gamma\gamma}=\frac{\eta l_1}{\pi R}\;2^{-\alpha-1}
 (1-\cos\theta_{\rm max})^{\alpha+2}\ep^\alpha,
\end{equation}
where
\begin{eqnarray}
\nonumber
 \eta(\alpha)=\frac{4}{\alpha+2}\int_1^\infty \ep_{\rm c}^{-2\alpha-1}\;
 \frac{\sigma_{\gamma\gamma}(\ep_{\rm c})}{\sigma_{\rm T}}\;{\rm d}\ep_{\rm c}
\end{eqnarray}
is the numerical factor already appearing in equation (3) and calculated 
analytically in Svensson (1987). 
Since the integral peaks at
$\ep_{\rm c}\sim 1$, the exact value of $\ep_{\rm max}$ is not important and we
set $\ep_{\rm max}=\infty$. Now from equations (3) and (8) we get
\begin{eqnarray}
\nonumber
  \frac{\kappa_{\gamma\gamma}}{\kappa_{\gamma\gamma}^{\rm isotrop}}=
  \sin^{2\alpha+4}\left(\frac{\theta_{\rm max}}{2}\right)=
  \frac{1}{2^{\alpha+2}}\left(1-\frac{z}{\sqrt{z^2+R^2}}\right)^{\alpha+2}.
\end{eqnarray}
%%%%%%%%%%%%%%%%%%%%%%%%%%%%%%%%%%%%%%%%%
\begin{figure}
\begin{center}
\leavevmode
\epsfxsize=8.4cm 
\epsfysize=8.2cm   
\epsfbox{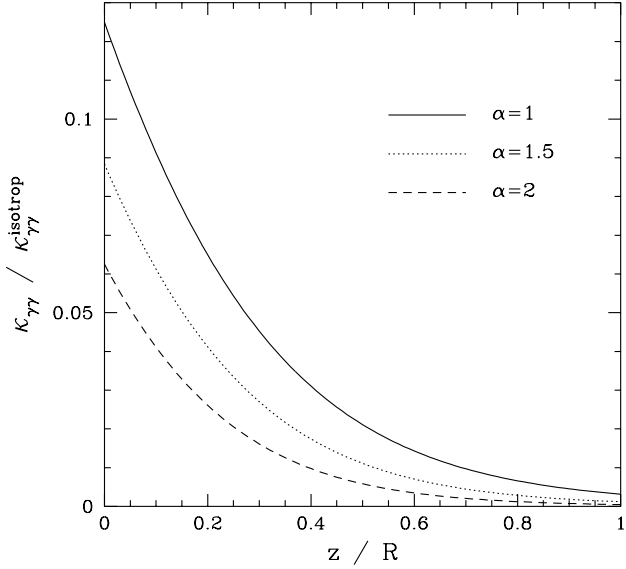}
\end{center}
\caption{The ratio of $\gga$ opacity at height $z$ above the disc to 
$\kappa_{\gamma\gamma}^{\rm isotrop}$ given by equation (3).
The opacity is calculated for $\gamma$-rays propagating along the disc axis.
$R$ is the disc radius, $\alpha$ is the spectral index of the radiation.}
\end{figure}
%%%%%%%%%%%%%%%%%%%%%%%%%%%%%%%%%%%%%%%%%
This function is shown in Fig. 1 for $\alpha=$1, 1.5, 2. The opacity above
the disc is strongly reduced 
because our photon moving along the disc axis does not 
encounter any radiation in the opposite direction which would dominate the
$\gga$ opacity in the isotropic case. 

It is interesting to compare the total optical depth
seen by a $\gamma$-photon emitted along the disc axis,
$\tau_{\gamma\gamma}=\int_0^\infty \kappa_{\gamma\gamma}{\rm d}z$, with 
$R\kappa_{\gamma\gamma}^{\rm isotrop}$ which can be (very roughly) associated
with the $\gamma-\gamma$ optical depth of a spherical source of the same 
intensity, $I_\ep$, and dimension, $R$. 
We get
\begin{eqnarray}
\nonumber
  \frac{\tau_{\gamma\gamma}}{R\kappa_{\gamma\gamma}^{\rm isotrop}}=
  \int_0^{\pi/2}\frac{\sin^{2\alpha+4}(\theta/2)}{\sin^2\theta}\;{\rm d}\theta.
\end{eqnarray}
The dependence of this ratio on $\alpha$ is shown in Fig. 2. E.g., for 
$\alpha=1$, $\tau_{\gamma\gamma}$ is $\approx$ 30 times smaller than 
$R\kappa_{\gamma\gamma}^{\rm isotrop}$.
This result combined with equation (4) shows that the absorption 
constraint  on $\gamma$-ray spectra is very sensitive to the source geometry
and it can be much weaker than the usually used estimate for the spherical
case. 

%%%%%%%%%%%%%%%%%%%%%%%%%%%%%%%%%%%%%%%%%%%
\begin{figure}
\begin{center}
\leavevmode
\epsfxsize=8.4cm 
\epsfysize=8.2cm 
\epsfbox{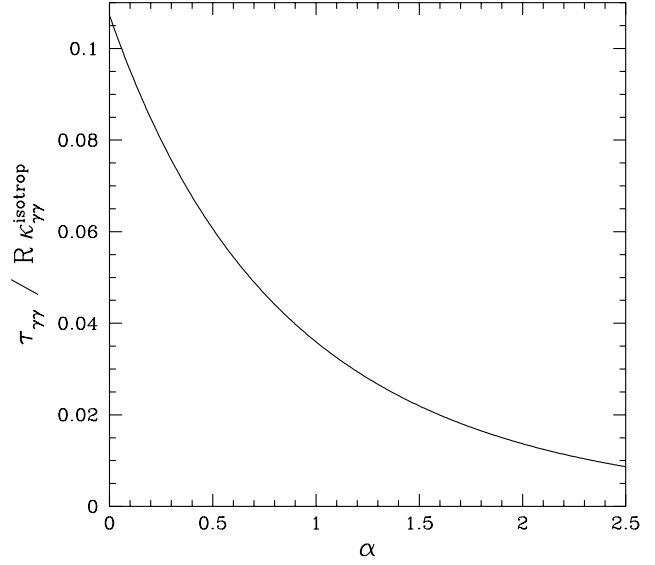}
\end{center}
\caption{The ratio of the $\gga$ optical depth seen by a gamma photon emitted
along the disc axis to $R\kappa_{\gamma\gamma}^{\rm isotrop}$;
$\alpha$ is the spectral index of the radiation.}
\end{figure}
%%%%%%%%%%%%%%%%%%%%%%%%%%%%%%%%%%%%%%%%%%%

\subsection{Dependence on inclination angle}

We now calculate the non-linear $\gamma$-ray transfer above the disc and
study the dependence of the observed absorption break on the disc 
inclination to the line of sight. The bulk of $\gga$ reactions occur close to 
the disc surface (see Fig. 1), and we solve a simplified transfer
problem in a one-dimensional approximation.
In this approximation, the 
radiation field is taken to be axisymmetric and uniform on any slice 
$z={\rm constant}$, and the effect of the finite size of the disc is treated 
for an optically thin atmosphere by truncating the radiation at angles
$\theta>\theta_{\rm max}(z)$. We divide the slab $z<R/2$ above the disc
into 25 horizontal layers, and assume that
the radiation escapes freely from the last (upper) layer.  
The intrinsic boundary condition is the isotropic power-law radiation 
given by equation (1). To get a stationary solution, $I_\ep(z,\theta)$,
we take an initially empty atmosphere and let the $\gamma$-rays propagate 
from the disc. We then follow the evolution of the radiation field in the 
computational slab until a stationary state is established. 
Further details of the numerical method are given in Appendix A.

Examples of spectra emerging at different angles from the disc with a modest
compactness $l_1=25$ and spectral index $\alpha=1.5$ are shown in Fig. 3. 
We included only the $\gga$ reaction in the transfer problem in order 
to see the efficiency of pure absorption. The other two 
non-linear effects, Compton scattering and annihilation emission of 
the produced pairs, become important when $l_1\simgt 30$ as the 
atmosphere becomes optically thick, see Section 3.1. 

A qualitative dependence on the inclination
angle $\theta$ is seen in Fig. 3: Absorption is more efficient 
at large inclinations. At large $\theta$, an emitted $\gamma$-photon
travels a long path at small heights where the $\gamma-\gamma$ opacity 
is large. Such a photon encounters especially high opacity because the 
threshold condition (5) is weaker at large $\theta$.
As a result, the escape probability is strongly reduced.

%%%%%%%%%%%%%%%%%%%%%%%%%%%%%%%%%%%%%%%%%%%%
\begin{figure}
\begin{center}
\leavevmode
\epsfxsize=8.4cm 
\epsfysize=8.2cm 
\epsfbox{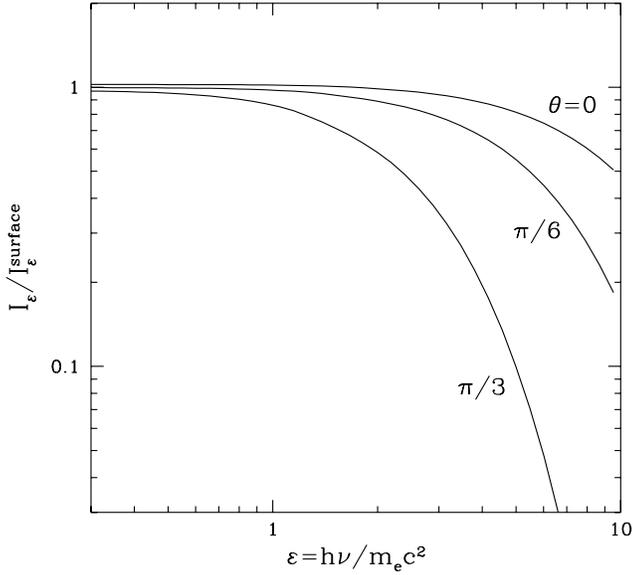}
\end{center}
\caption{The $\gga$ absorption observed at inclination 
angles $\theta=0,\;\pi/6,\;\pi/3$ from a disc with compactness $l_1=25$ and 
spectral index $\alpha=1.5$. $I_\ep^{\rm surface}$
is the radiation intensity at the disc surface given by equation (1). 
$I_\ep$ is the intensity of the emerging radiation found by numerical 
simulation of the $\gamma$-ray transfer above the disc. Compton scattering 
and annihilation emission in the $e^\pm$ atmosphere are neglected.}
\end{figure}
%%%%%%%%%%%%%%%%%%%%%%%%%%%%%%%%%%%%%%%%%%%%

%###########################################################################

\section{Optically thin pair wind}

Each $\gga$ interaction produces one $e^\pm$ pair. Therefore,
the solution to the transfer problem in Section 2.3
automatically gives the pair injection rate, $\dot{n}_+^{\gamma\gamma}$, 
as a function 
of the height $z$ above the disc. The bulk of pairs are injected close to the 
disc surface and are immediately cooled by the soft radiation down to 
the Compton temperature of the radiation field, $kT_{\rm C}\sim 1-10$ keV.
The time-scale for Compton cooling is given by
\begin{equation}
  t_{\rm C}=\frac{3m_{\rm e}c}{8\sigma_{\rm T}w}=
            \frac{3}{16}\frac{m_{\rm e}}{m_{\rm p}}\frac{R}{c}
   \left(\frac{L}{L_{\rm E}}\right)^{-1}\left(\frac{R}{r_{\rm g}}\right),
\end{equation}
where $w=L/\pi R^2c$ is the radiation energy density,
$L$ is the disc luminosity. Here only luminous discs are considered, and 
$t_{\rm C}$ is shorter than the time-scale for pair escape, 
$t_{\rm esc}\sim R/c$.
On the time-scale $t_{\rm C}$, pairs are also accelerated by the pressure of 
soft radiation. 
The $e^\pm$ plasma is light and the gravitational force can be neglected as 
compared to the radiative force when $L>(m_{\rm e}/m_{\rm p})L_{\rm E}$. 
Then pairs acquire an equilibrium bulk velocity for which the radiation 
pressure is balanced by the radiation drag. Near the disc surface, where the 
radiation field is semi-isotropic, the equilibrium velocity is about $0.45 c$ 
(Icke 1989; see also Section 4.2).

\subsection{Luminosity in pair rest mass}

In an optically thin wind, pairs escape before they can annihilate 
since $t_{\rm esc}$ is less than the time-scale for annihilation,
$t_{\rm ann}\sim (n_+\sigma_{\rm T}c)^{-1}$. Then the flux of escaping
pairs equals the column rate of pair production above the disc.
The corresponding flux of $e^\pm$ rest-mass energy equals
\begin{eqnarray}
\nonumber
   F_\pm=2m_{\rm e}c^2\int_0^{R/2}\dot{n}_+^{\gamma\gamma}(z){\rm d}z.
\end{eqnarray}
The total rest-mass energy carried away by the $e^\pm$ wind can be estimated as 
\begin{eqnarray}
\nonumber
  L_\pm= 2 \pi R^2 F_\pm.
\end{eqnarray}

In the limit of small compactness, $\gamma$-rays propagate almost
freely in the atmosphere, and only a small fraction of $L_1$ is
transformed into $L_\pm$. This fraction is proportional
to the $\gga$ optical depth which, in turn, is proportional to $l_1$. 
Therefore, in the limit of small compactness $L_\pm/L_1\propto l_1$.
As $l_1$ increases, $\gga$ absorption significantly depletes MeV photons, 
which reduces the $\gga$ opacity. The transfer becomes non-linear
and $L_\pm/L_1$ saturates at $\simlt 20$ \%. 
Numerically calculated values of $L_\pm/L_1$ for several compactnesses $l_1$ 
are given in Table 1. In the calculations, a spectral index $\alpha=1.5$ was 
assumed.

%%%%%%%%%%%%%%%%%%%%%%%%%%%%%%%%%%%%%%%%%%%%%%%%%%
\begin{table}
\caption[ ]{The efficiency of $\gamma$-ray transformation into an $e^\pm$-wind}
\begin{center}
{\normalsize
\begin{tabular}{ccccccccc}\hline 
$l_1$       &  5    &  10  &  15  &  20  &  25  &  30  \\ [0.8ex]
$L_\pm/L_1$ & 0.04  & 0.07 & 0.09 & 0.10 & 0.11 & 0.12 \\ [0.8ex]
\hline
\end{tabular}
}
\end{center}
\end{table}
%%%%%%%%%%%%%%%%%%%%%%%%%%%%%%%%%%%%%%%%%%%%%%%%%%

Pairs escape with velocities $\sim c/2$, and the density of the atmosphere
can be estimated as $n_+\sim L_\pm/2\pi R^2m_{\rm e}c^3$. The typical 
scattering optical depth of the $e^\pm$-cloud around the disc is then given by
\begin{equation}
\tau_{\rm T}\sim 2n_+\sigma_{\rm T} R\sim \frac{l_1}{\pi}
        \left(\frac{L_\pm}{L_1}\right).
\end{equation}
Using Table 1, one can see that the atmosphere
becomes optically thick when $l_1\sim 30$. 

In an optically thick wind, 
$t_{\rm ann}$ would become comparable to $t_{\rm esc}$. Then $L_\pm$
is limited by annihilation which depletes the particle population 
in the wind. The density of escaping pairs is controlled by the condition
$t_{\rm ann}\sim t_{\rm esc}$ which gives $n_+\sim (\sigma_{\rm T}R)^{-1}$
and the upper limit for $L_\pm$ (Phinney 1983; Guilbert \& Stepney 1985;
Ghisellini et al. 1992), 
\begin{equation}
  L_\pm^{\rm max}\sim \frac{2\pi m_{\rm e}c^3 R}{\sigma_{\rm T}},
\end{equation}
which corresponds to compactness $l_\pm\sim 2\pi$.

\subsection{Annihilation emission}

Consider a plasma volume ${\rm d}V$ moving with velocity $\bv=\bbeta c$.
In the comoving frame, the number of annihilation photons emitted by this 
volume into solid angle
${\rm d}\tilde{\Omega}$ during time ${\rm d}\tilde{t}$ equals
$$
  {\rm d}N=\frac{3}{16\pi}\,c\sigma_{\rm T} 
  \tilde{n}_{+}^2 {\rm d}\tilde{V}{\rm d}\tilde{t} 
   {\rm d}\tilde{\Omega},
$$
where $\tilde{n}_+=n_+/\gamma$ is the pair density in the comoving frame,
${\rm d}\tilde{V}=\gamma {\rm d}V$, and $\gamma=(1-\beta^2)^{-1/2}$. 
In lab frame, these photons propagate within a solid angle 
${\rm d}\Omega={\rm d}\tilde{\Omega}\ep_{\rm ann}^{-2}$,
where $\ep_{\rm ann}=\gamma^{-1}(1-\bbeta\cdot{\bf\Omega})^{-1}$
is the photon energy in lab frame. Taking into account the invariance 
of four-volume, ${\rm d}\tilde{V}{\rm d}\tilde{t}={\rm d}V{\rm d}t$, we get 
the annihilation power emitted in direction ${\bf\Omega}$ by a unit volume of 
a stationary wind, 
\begin{eqnarray}
\nonumber
Q_{\rm ann}({\bf\Omega})=\frac{\ep_{\rm ann}m_{\rm e}c^2 {\rm d}N}
                              {{\rm d}V{\rm d}t{\rm d}\Omega}
         =\frac{3m_{\rm e}c^3\sigma_{\rm T}n_+^2}
         {16\pi\gamma^5(1-\bbeta\cdot{\bf\Omega})^3}.
\end{eqnarray}
Note that the fraction of photons emitted within a solid angle 
${\rm d}\Omega$ equals ${\rm d}\Omega\ep_{\rm ann}^2/4\pi$. 
The average energy of the annihilation photons equals $\gamma m_{\rm e}c^2$.

A mildly relativistic bulk motion makes the wind emission considerably 
different from that of a cold source at rest:

i) $Q_{\rm ann}$ is significantly anisotropic even for a modest $\beta$.
E.g., for $\beta=0.5$ the emission in the flow direction is 
8 times larger than in the perpendicular direction.

ii) The annihilation photons are blueshifted in the lab frame by 
a factor of $[\gamma(1-\bbeta\cdot{\bf\Omega})]^{-1}$. For a 
face-on-oriented observer and $\beta=0.5$ the blueshift equals 
$\sqrt{3}$.

iii) The width of the annihilation line is determined by the velocity gradients
in the wind rather than by the thermal motions. The bulk velocities are 
relativistic, so the line must be broad.

The total power of the annihilation emission can be roughly estimated as
\begin{eqnarray}
\nonumber
  L_{\rm ann}\sim \left(\frac{t_{\rm esc}}{t_{\rm ann}}\right)L_\pm
             \sim\tau_{\rm T} L_\pm,
\end{eqnarray}
where $\tau_{\rm T}$ is the typical scattering optical depth of the atmosphere, 
see equation (10). In the limit of small compactnesses, $L_{\rm ann}/L_1$ falls
off $\propto l_1^3$. Then annihilation of pairs produced inside the source is 
likely to dominate, and a narrow line at 511 keV might be expected. In the case
of large compactnesses, where the 
atmosphere becomes optically thick, the line produced inside the source must 
be scattered above the disc. The observed annihilation emission is then 
dominated by the $e^\pm$ outflow which produces a broad blueshifted line 
as discussed in next Section.

%############################################################################

\section{Optically thick outflow}

We now address the formation of an optically thick outflow by calculating
the $\gamma$-ray transfer above a disc with a compactness $l_1>30$.
The main conversion of $\gamma$-rays into $e^\pm$ pairs 
occurs close to the disc surface, at heights $z\ll R$ (see Section 4.1). 
We therefore consider a simplified one-dimensional transfer problem, in which
pair production is calculated in a slab $0<z<R/2$. We thus 
neglect the boundary effects connected with the finite size of the disc.
The radiation is assumed to escape freely at the outer boundary $z=R/2$.
Further calculations verify that this is a reasonable first approximation.

A magnetic field may affect the outflow dynamics
if the magnetic pressure exceeds the radiation pressure. The configuration 
of the accretion disc magnetosphere is, however, quite uncertain. 
We assume here that the magnetic field does not prevent the vertical $e^\pm$ 
outflow.

In an optically thick outflow 
$t_{\rm ann}/t_{\rm esc}\sim (n_+\sigma_{\rm T}R)^{-1}< 1$, i.e.,
pairs annihilate before they can escape.
The bulk of $e^\pm$ form a thermal population at the
Compton temperature, $kT_{\rm C}\sim 1-10$ keV, and only a small 
fraction form a non-thermal tail.
The tail consists of those pairs which are braking after their creation 
towards thermal energies due to inverse Compton (IC) cooling by soft radiation
and Coulomb collisions with the thermal $e^\pm$ plasma.
The stationary energy distribution of the braking particles peaks at 
semi-relativistic energies (Beloborodov \& Illarionov 1995).
The density of non-thermal pairs $n_+^{\rm nth}\approx \dot{n}_+^{\gamma\gamma}
t_{\rm C}$ is a small fraction, $\sim t_{\rm C}/t_{\rm ann}$, of the 
thermal population, $n_+$.

The pairs are created in the interactions between MeV photons 
and they typically have initial Lorentz factors $\gamma\sim 2 - 3$. 
IC photons produced in the slab therefore have modest energies
$\ep_{\rm IC}\sim \gamma^2\ep_{\rm s}\simlt 50$ keV, where $\ep_{\rm s}$ is 
the energy of a soft photon before upscattering, typically in the range 
1 eV -- 10 keV. (The non-thermal component is optically thin and 
repeated upscatterings are negligible.) 
It means that the kinetic energy of created pairs is just absorbed
by the soft component of radiation and 
a negligibly small fraction is converted into hard radiation. 
We therefore neglect the presence of non-thermal pairs 
in the calculations of hard X/$\gamma$-ray transfer and assume
that the plasma is composed of the cold thermal pairs only.
Then there are only two sources of hard photons: the primary source below the 
computational slab and the annihilation emission of pairs produced in the slab.

In calculations of Compton scattering by thermal $e^\pm$ we neglect their slow 
thermal motions and account for the fast bulk motion
which strongly affects the scattering process. The plasma has a bulk velocity 
near the equilibrium value $v\sim 0.5c$ determined by the soft radiation 
which dominates the energy density of the atmosphere. 
The equilibrium velocity can be found as 
a function of optical depth independently of the atmosphere density profile
(see Section 4.2). The velocity varies from 
$0.3 c$ to $0.7c$. As a first step, we adopt $\beta=v/c=0.5$ in the simulations
of the $\gamma$-ray transfer.

\subsection{Numerical model of outflow formation}

We have calculated several transfer models assuming
an isotropic source (1) below the computational slab. 
Details of the code are given in Appendix A.

%%%%%%%%%%%%%%%%%%%%%%%%%%%%%%%%%%%%%%%%%%%%
\begin{figure}
\begin{center}
\leavevmode
\epsfxsize=8.4cm 
\epsfysize=8.2cm 
\epsfbox{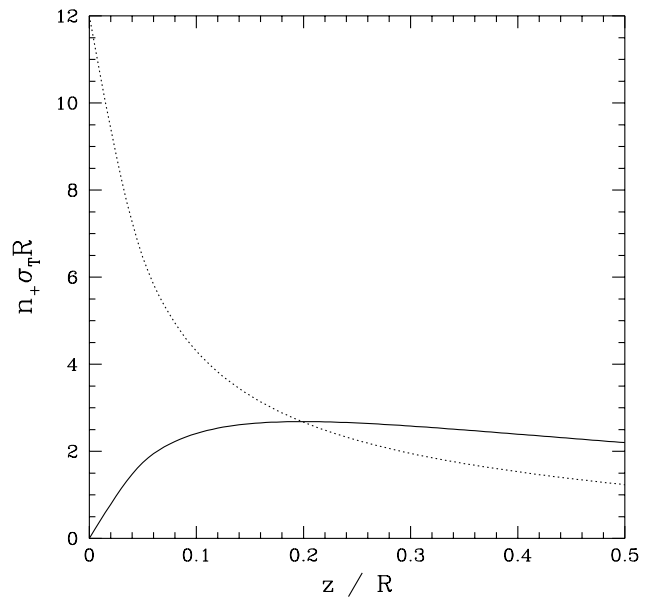}
\end{center}
\caption{The density profile of the $e^\pm$ outflow created above a disc 
with compactness $l_1=150$ and spectral index $\alpha=1.5$ (solid curve). 
The dotted curve shows the density $n_+^*$ which would correspond to the 
local balance between pair production and annihilation.}
\end{figure}
%%%%%%%%%%%%%%%%%%%%%%%%%%%%%%%%%%%%%%%%%%%%

We here consider the case where there is no $e^\pm$ flux from the 
$\gamma$-ray source and set $n_+=0$ at $z=0$. Then the outflow is created 
above the source.  
An example of the density profile obtained for the atmosphere above a disc with
a compactness $l_1=150$ and a spectral index $\alpha=1.5$ is shown in Fig. 4. 
The equilibrium density, $n_+^*(z)$, would correspond to local balance
between pair production and annihilation, i.e., 
$\dot{n}_+^{\gamma\gamma}=(3/8)(1-\beta^2)\sigma_{\rm T}c{n_+^*}^2$.
The pair density, $n_+$, strongly deviates from $n_+^*$ as a result of pair 
advection along the $z$-axis. The $e^\pm$ transfer equation in the stationary 
case can be written as 
\begin{eqnarray}
 \beta\frac{{\rm d}n_+}{{\rm d}z}=\frac{3}{8}(1-\beta^2)\sigma_{\rm T}
 \left[{n_+^*}^2-n_+^2\right].
\nonumber
\end{eqnarray}
The first and second terms in the square braquets stand for $e^\pm$ 
production and annihilation respectively. At large heights, advection 
results in that $n_+>n_+^*$ and pair production becomes unimportant:
the outflow is dominated by pairs advected from the lower layers. The evolution
of the pair density with height is then governed mainly by annihilation. 

The steep decrease in $n_+^*$ shows that the rate of pair injection, 
$\dot{n}_+^{\gamma\gamma}\propto {n_+^*}^2$, steeply 
decreases with height as a result of strong degradation of the 
%%%%%%%%%%%%
$\gamma$-rays\footnote{Note that the radiation quickly becomes 
anisotropic in the low layers because of preferential depletion 
of $\gamma$-rays at large $\theta$. The boundary effects
are therefore reduced and the slab geometry may be a good approximation,
in contrast to the optically thin case where the finiteness of the disc radius 
was important and we truncated the radiation at $\theta>{\rm arctan}(R/z)$.}. 
%%%%%%%%%%%%
Roughly, the outflow is created in the ``injection zone'', $z/R\simlt 0.2$,
where $n_+<n_+^*$. The created pair plasma then outflows in the ``annihilation
zone'', $z\simgt 0.2$, where $n_+>n_+^*$.

The intensity of the outgoing $\gamma$-rays is suppressed to such an extent 
that the equilibrium pair density becomes $n_+^*\sim(\sigma_{\rm T}R)^{-1}$
which is also the typical density of the ``false photosphere''.
At $z\sim R$ we get 
$n_+\sim (\sigma_{\rm T}R)^{-1}$, i.e., we are already in the transition zone 
between the optically thick $e^\pm$ envelope and an optically
thin wind from the envelope. At $z>R$ the flow is no longer plane-parallel
and the pair density is strongly reduced.
Our code is not able to follow the transition
to the optically thin wind. One can only estimate the 
flux of escaping pairs as $\sim c/\sigma_{\rm T}R$. The emerging luminosity
in $e^\pm$ rest mass is given by equation (11).

We now briefly discuss the annihilation emission of the outflow shown in 
Fig. 4. Only photons emitted perpendicular to the disc have a 
significant chance ($\sim 0.5$) to escape the optically thick layers,
and the bulk of photons emitted at large angles will be 
downscattered or absorbed. The escaping photons will be detected by a 
distant observer as an annihilation line. In the numerical simulation we 
obtained a line in the emerging spectrum located at 
$\ep_{\rm ann}=[\gamma(1-\beta\cos\theta)]^{-1}$ where $\beta=0.5$ is 
the atmospheric velocity and $\theta$ is inclination angle. With increasing
inclination, the line flux as well as the continuum $\gamma$-ray emission
decreases by 10 -- 30 times. The equivalent width of the line is 
$\sim\ep_{\rm ann}/2$ at all inclinations. A model taking into account the 
velocity
gradient in the optically thick layers (see Section 4.2) would yield a broad 
annihilation bump of approximately the same equivalent width.

\subsection{Collimation of soft radiation}

The $\gamma$-ray flux converted into pairs above the disc is much smaller than 
the flux of soft radiation. The latter, being approximately equal to the total 
energy flux from the disc, is constant in the optically thick $e^\pm$ outflow. 
However, scattering in the outflow changes the angular distribution of the
soft radiation.

The scattering mainly occurs in the optically thick layers, $z<R$, where the 
outflow is approximately plane-parallel. We consider the simplified 
one-dimensional problem of radiative transfer in a slab of vertically 
outflowing pair plasma atop an isotropic source of soft radiation.
Each electron/positron moves in Compton equilibrium during its life-time in the 
$e^\pm$ envelope. The equilibrium velocity at each height is 
determined by the local angular distribution of the frequency-integrated 
intensity of the radiation, $I$.
%We now address the transfer of soft radiation in an $e^\pm$ outflow with 
%a given column density, assuming that the scattering pairs move 
%in Compton equilibrium with the radiation. 

In the plane-parallel geometry, the intensity is a function of
the angle $\theta$ between the ray and the $z$-axis, and of the optical depth
in the $e^\pm$ envelope, ${\rm d}\tau=-2n_+\sigma_{\rm T}{\rm d}z$. 
The only parameter of the problem is the column density of the envelope, 
$\tau_0/\sigma_{\rm T}$. Here we neglect the polarization of 
radiation due to scattering (polarization is considered in Beloborodov 1998). 
The transfer equation then reads
\begin{eqnarray}
\nonumber
   \frac{1}{2n_+\sigma_{\rm T}c}\frac{\partial I}{\partial t}-
   \cos\theta\frac{\partial I}{\partial\tau}= (1-\beta\cos\theta)(S-I), \\
   0<\tau<\tau_0,
\end{eqnarray}
The radiation scattered in direction ${\bf\Omega}$ is represented by the 
source function, 
\begin{equation}
  S({\bf\Omega})=\frac{1}{\sigma_{\rm T}(1-\bbeta\cdot{\bf\Omega})}
                 \int I({\bf n})\,
                 \frac{{\rm d}\sigma}{{\rm d}\Omega}({\bf n})\,
                 \frac{\ep^\prime}{\ep}\;{\rm d}{\bf n},
\end{equation}
where ${\bf n}$ is the direction of a photon before scattering and 
${\rm d}\sigma/{\rm d}\Omega$ is the differential cross section for Thomson 
scattering.
$\ep$ and $\ep^\prime$ are the photon energies before and 
after the scattering respectively, and they are related by the equation
\begin{equation}
  \ep^\prime(1-\bbeta\cdot{\bf\Omega})=\ep(1-\bbeta\cdot{\bf n}),
\end{equation}
which expresses the fact that the photon energy does not change in 
the comoving frame.

The equilibrium velocity $\beta$ is determined at each height by the equation
\begin{equation}
   (1+\beta^2)I_1-\beta (I_0+I_2)=0,
\end{equation}
where
\begin{eqnarray}
\nonumber
   I_m=\frac{1}{4\pi}\int I(\theta)\cos^m\theta\; {\rm d}{\bf\Omega}.
\end{eqnarray}
The equilibrium equation expresses the condition that the net flux of
radiation in the comoving frame vanishes (see, e.g., Sikora \& Wilson 1981),
and in the lab frame it means that 
the radiation pressure is balanced by the radiation drag.
Pairs keep their velocity near the equilibrium value which varies with height. 
The time-scale for relaxation to the equilibrium $\sim t_{\rm C}$ is the 
shortest time-scale in the problem, therefore, the equilibrium velocity is a 
strong attractor in phase space, and we neglect deviations of $\beta$ from that 
given by equation (15). The equations (12) and (15) then form a closed set of 
equations.  

We are looking for a stationary solution, $\partial I/ \partial t=0$.
The first and second moments of the radiation field are integrals of the 
problem, $I_1(\tau)={\rm constant}$ and $I_2(\tau)={\rm constant}$,
(in contrast to the classical problem of radiative transfer in an electron 
medium at rest, which has only $I_1(\tau)={\rm constant}$). The constancy of 
$I_2$ can easily
be checked by combining the first moment of equation (12) with 
the condition (15). $I_2(\tau)={\rm constant}$ is just another way
to express the equilibrium condition: $4\pi I_2/c$ equals the radiation 
pressure in the vertical direction, and its gradient vanishes in the 
equilibrium $e^\pm$ envelope.

Another special feature of the equilibrium radiative transfer is 
that the radiation density inside an $e^\pm$ envelope cannot be strongly 
enhanced regardless the optical depth of the envelope. 
The trapped radiation  
is advected out of the optically thick layers with a velocity $\sim c/2$,
and the presence of the envelope around the disc weakly
affects the radiation density, 
as can also be derived formally using $I_2(\tau)={\rm constant}$. 
The main
impact of $e^\pm$ is on the angular distribution of radiation.  

Here we adopt a simplified model for Thomson scattering.
We will assume that the scattered radiation is isotropic in the comoving frame,
i.e., a fraction ${\rm d}\tilde\Omega/4\pi$ of the scattered photons have 
directions within the solid angle ${\rm d}\tilde\Omega$.
The solid angle corresponding to ${\rm d}\tilde\Omega$ in lab frame equals 
${\rm d}\Omega= \gamma^2(1-\bbeta\cdot{\bf\Omega})^2{\rm d}\tilde\Omega$
(see, e.g., Rybicki \& Lightman 1979). 
The total scattering cross section equals 
$\sigma=\sigma_{\rm T}(1-\bbeta\cdot{\bf n})$,
and the differential cross section 
for the scattering in the lab frame is given by
\begin{eqnarray}
\nonumber
  \frac{{\rm d}\sigma}{{\rm d}\Omega}=
  \frac{\sigma_{\rm T}}{4\pi}\frac{1-\bbeta\cdot{\bf n}}
                            {\gamma^2(1-\bbeta\cdot{\bf\Omega})^2}.
\end{eqnarray}
Equations (13) and (14) then give the source function
\begin{eqnarray}
\nonumber
  S(\theta)=\frac{I_0-2\beta I_1+\beta^2 I_2}{\gamma^2 (1-\beta\cos\theta)^4}.
\end{eqnarray}
Using the equilibrium condition (15), we express $I_2$ in terms of
$I_0$, $I_1$, and $\beta$, and rewrite $S(\theta)$ as
\begin{equation}
  S(\theta)=\frac{I_0-\beta I_1}{\gamma^4 (1-\beta\cos\theta)^4}.
\end{equation}

%%%%%%%%%%%%%%%%%%%%%%%%%%%%%%%%%%%%%%%%%%%%%%%%%%%
\begin{figure*}
\begin{center}
\leavevmode
\epsfxsize=8.4cm 
\epsfysize=8.2cm 
\epsfbox{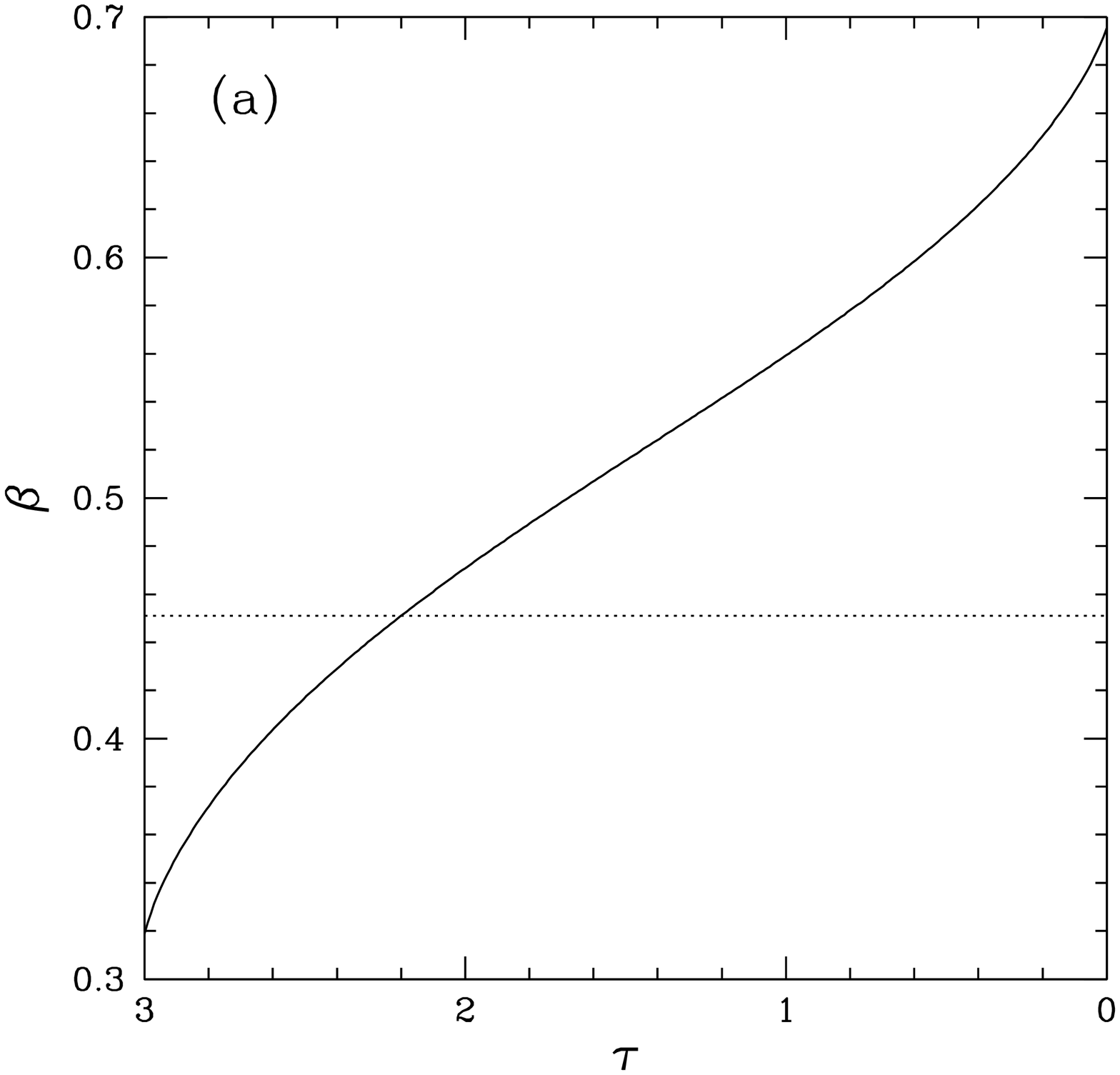}
\epsfxsize=8.4cm 
\epsfysize=8.2cm 
\epsfbox{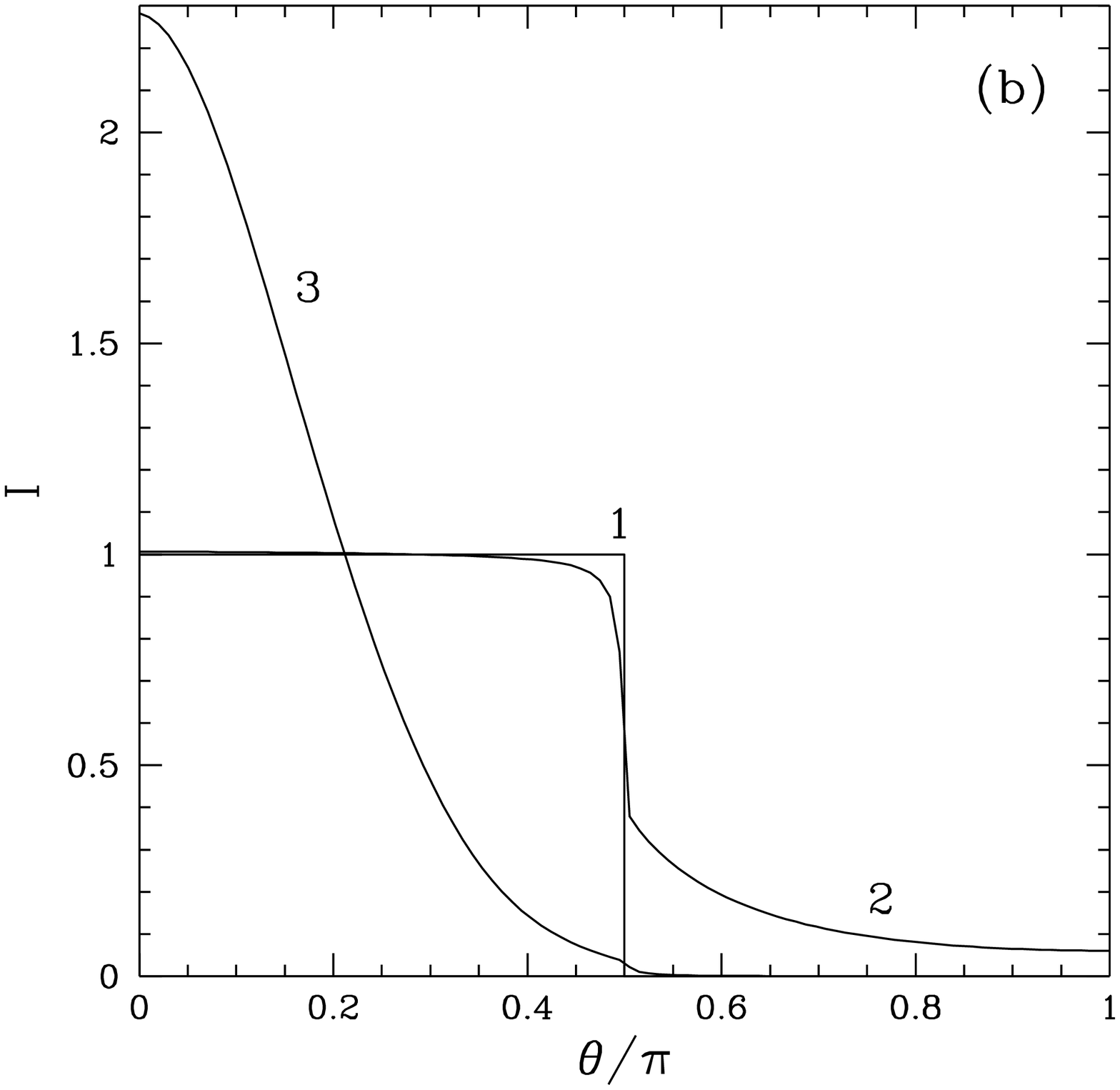}
\end{center}
\caption{The stationary solution for the soft radiation transfer in a
plane-parallel $e^\pm$ atmosphere with column density $3\sigma_{\rm T}^{-1}$:
(a) Velocity profile. Parameter $\tau$ is defined as
${\rm d}\tau=-2n_+\sigma_{\rm T}{\rm d}z$. The dotted line displays the 
equilibrium velocity in a semi-isotropic radiation field. 
(b) The angular dependence of the soft radiation intensity: 
1 -- as emitted by the disc, 2 -- at the base of the 
atmosphere ($\tau=3$), 3 -- as emerging from the atmosphere ($\tau=0$). 
The intensity is normalized so that $I({\bf\Omega})=1$ for the radiation 
emitted by the disc.} 
\end{figure*}
%%%%%%%%%%%%%%%%%%%%%%%%%%%%%%%%%%%%%%%%%%%%%%%%%%%

We solve the equations (12) and (15) with the source function (16) numerically.
At the inner boundary 
($\tau=\tau_0$), we set a source of isotropic soft radiation, which absorbs  
photons impinging back on the disc (the results are essentially the same if 
we assume reflection at the inner boundary). At the outer boundary ($\tau=0$), 
we assume free escape of the radiation. 
As initial conditions we assume that a semi-isotropic radiation is filling 
the atmosphere
at all heights (curve 1 in Fig. 5b). This would be a stationary solution if 
the atmosphere were optically thin (the corresponding equilibrium velocity
$c(4-\sqrt{7})/3\approx 0.45 c$ is shown by the dotted line in Fig. 5a). 
The initial state evolves due to Thomson scattering in the atmosphere. 
The stationary solution $I(\tau,\theta)$,
$\beta(\tau)$ that we are looking for depends only on the parameter $\tau_0$
and does not depend on the density profile, $n_+(z)$.
We therefore choose the simplest density profile, $n_+(z)={\rm constant}$. 
In the calculations, we use a grid which is homogeneous in the $z$-- and 
$\theta$--direction. The number of grid points is $N_z\times 
N_\theta=300\times 100$. 
The chosen time step equals $\Delta t=0.3\tau_0/2n_+\sigma_{\rm T}cN_z$. 

The resulting stationary solution for $\tau_0=3$ is presented in Fig. 5a,b. 
In the lowest layers of the slab, the equilibrium velocity is less than $0.45c$
since, besides the semi-isotropic disc radiation, there is some flux in the 
opposite direction as a result of the radiation being 
backscattered by the outflow (see Fig. 5b, curve 2). 
In higher layers, the radiation gets collimated by scattering on 
the moving pairs, and the equilibrium velocity increases to $\sim 0.7 c$.
The angular distribution of the outgoing radiation is shown by curve 3 
in Fig. 5b. The distribution does not depend on $\tau_0$ in the optically
thick limit, and therefore does not depend on the compactness of the primary
$\gamma$-ray source.
The emerging radiation is significantly anisotropic, e.g., 
the intensity at $\theta=\pi/3$ is $\sim 7$ times smaller than that at 
$\theta=0$.

\subsection{Acceleration of escaping pairs}

The escaping pairs form an optically thin wind.
At large heights, $z\gg R$, the angular size of the central luminous disc 
$R\sim r_{\rm g}$ 
diminishes and the escaping pairs are accelerated. Their equilibrium 
Lorentz factor is determined by the balance between pressure of the central 
radiation and drag by the radiation from the outer parts of the 
disc (Phinney 1987). With increasing height the density of the central 
radiation 
is reduced $\propto z^{-2}$ and redshifted in the comoving frame as 
$\gamma^{-2}$. The density of the diffuse radiation from the outer parts of 
the disc scales $\propto z^{-3}$, but it is blueshifted as $\gamma^2$. 
The resulting equilibrium Lorentz factor 
$\gamma_{\rm eq}\sim (z/r_{\rm g})^{1/4}$. At large distances from the black 
hole, the pair Lorentz factor falls below the equilibrium value because of 
radiation dilution, and saturates at 
$\gamma_\infty\sim (m_{\rm p}L/m_{\rm e}L_{\rm E})^{1/7}$, where $L$ is
the disc luminosity, $L_{\rm E}$ is the Eddington limit. 

The collimation of the central radiation may increase $\gamma_\infty$.
The relevant collimation parameter, $\zeta$, 
equals the ratio of the collimated intensity at $\theta=0$ to the intensity of
isotropic radiation of the same net flux. In our case $\zeta\approx 2.3$,
see Fig. 5b. 
Accounting for the collimation, the equilibrium Lorentz factor can be estimated
as $\gamma_{\rm eq}\sim \zeta^{1/4}(z/r_{\rm g})^{1/4}$. 
At some height $z_*$, $\gamma$ falls below $\gamma_{\rm eq}$.
At heights $z\gg z_*$, 
the drag is negligible and the acceleration is determined by the central
radiation flux $F=\zeta L/4\pi z^2$, 
\begin{eqnarray}
\nonumber
   \dot{\gamma}\approx \frac{\sigma_{\rm T}F}{4\gamma^2m_{\rm e}c^2}.
\end{eqnarray}
Substituting $\dot{\gamma}\approx c({\rm d}\gamma/{\rm d}z)$ and
integrating this equation gives the asymptotic solution 
\begin{eqnarray}
\nonumber
  \gamma^3(z)= \gamma_\infty^3
  -\frac{3}{8}\frac{m_{\rm p}}{m_{\rm e}}\frac{\zeta L}{L_{\rm E}}
   \frac{r_{\rm g}}{z}, \qquad z\gg z_*.
\end{eqnarray}
The falling of $\gamma$ below $\gamma_{\rm eq}$
and transition to the regime $\gamma\ll\gamma_{\rm eq}$ occurs  when
the acceleration time-scale, $\gamma/\dot{\gamma}$,
equals the escape time-scale, $z/c$. At $z=z_*$ one therefore has the condition
$8\gamma^3(z/r_{\rm g})\approx (m_{\rm p}/m_{\rm e})(\zeta L/L_{\rm E})$.
By matching the solutions for the two regimes $\gamma=\gamma_{\rm eq}$ and 
$\gamma\ll\gamma_{\rm eq}$ at $z=z_*$, one can estimate $z_*$ and then 
$\gamma_\infty$, 
\begin{equation}
  \gamma_\infty\approx\zeta^{2/7}\left(\frac{m_{\rm p}}{m_{\rm e}}
              \frac{L}{L_{\rm E}}\right)^{1/7}.
\end{equation}
The weak dependence on $\zeta$ shows that $\gamma_\infty$ is not crucially
affected by the collimation of the central radiation.
To obtain large Lorentz factors $\sim 10-20$ observed in AGN 
jets, a collimation factor $\zeta\sim 10^2$ would be required.

%#########################################################################

\section{Conclusions}

MeV photons are strongly absorbed above a compact disk-like source
and the observed $\gamma$-ray luminosity is suprressed, 
especially if the disc is viewed at large inclinations. 
The pair plasma created above the disc flows away in Compton equilibrium 
with the soft radiation field.
The optical depth of the outflow depends on the $\gamma$-ray compactness, 
$l_1$, (see Section 2.1 for the definition of $l_1$). 
The outflow is optically thin if $l_1<30$ and becomes optically thick 
if $l_1\simgt 30$.
In the optically thin case, pairs escape without annihilation and the 
emerging luminosity in pair rest mass corresponds to the number of
$\gamma$-photons absorbed above the disc.

The optically thick outflow can be described as an envelope surrounding the 
disc where the created pairs have a life-time less than the escape time-scale. 
During their life-time pairs move in equilibrium with the soft radiation field
with a velocity $\sim 0.5 c$. 
Detailed calculations of the optically thick outflow show that its 
velocity increases from $\approx 0.3 c$
at the base to $\approx 0.7 c$ at the photosphere. 
In the photospheric layers, $z\sim R$, the pair density reduces to 
$n_+\sim (\sigma_{\rm T}R)^{-1}$, $R$ being the
radius of the $\gamma$-ray emitting disc. 
A significant fraction of pairs created in these
layers escape to produce a luminosity in pair rest mass
$L_\pm \sim 2\pi m_{\rm e}c^3R/\sigma_{\rm T}$.

The condition for formation of an optically thick $e^\pm$ outflow is likely
to be fulfilled in the brightest objects, such as quasars, especially if the 
black hole has a large spin. The bulk of the energy is then released in a 
very compact region, $R\approx r_g$, and the outflow becomes optically 
thick if the luminosity above 511 keV exceeds a small fraction,
$\sim 3\times 10^{-3}$, of the Eddington luminosity. 
The optically thick outflow will strongly affect the observed radiation:
 
i) The luminosity of the central region becomes anisotropic
as a result of scattering in the outflow.
The apparent bolometric luminosity of the disc is then suppressed if the disc 
is viewed at large inclinations. 
 
ii) The outflow obscures the central region of the accretion
disc and inhibits observation of reflection features, in particular the
Fe K$\alpha$ line. Reflected X-rays may be observed from the outer parts of
the disc which is illuminated by the central source. However, because of the 
strong anisotropy of the source, the reflection features should be weak.
This may explain the lack of Fe K$\alpha$ lines in bright radio-quiet quasars 
compared with lower luminosity Seyfert AGNs (Reeves et al. 1997).
 
iii) The expected annihilation line is blueshifted and broad due to
the velocity gradient in the outflow. The line increases the
continuum $\gamma$-ray flux by a factor of $\sim 2$ at $0.5 - 1$ MeV,
in contrast to the narrow annihilation feature at 511 keV usually expected
from luminous $\gamma$-ray sources. 

Even an outflow of a modest optical depth, $\sim 0.1$, may affect the observed
radiation by changing its polarization (Beloborodov 1998):
the polarization vector becomes parallel to the disc axis, in agreement with
optical observations of non-blazar AGNs (Antonucci 1992). 

Our calculations of the pair production around the accretion disc are limited 
to the one-dimensional slab approximation. Simulations of the outflow formation
in cylindrical geometry may provide new insights into compact $\gamma$-ray 
sources.
The $\gamma$-ray emitting region of the disc has a finite radius and at large 
inclinations the slab approximation is not adequate as the bulk of observed 
hard radiation then comes from the boundary of the $\gamma$-ray emitting 
region. Simulations in cylindrical geometry would also allow the study of the 
interesting case where a source of GeV radiation is geometrically 
separated from the source of soft X-rays, being located in the centre of the 
surrounding X-ray ring (Illarionov \& Krolik 1996). In such a geometry, the 
$e^\pm$ outflow would be created in a cylindrical wall between the two sources.

\section*{Acknowledgments}

I am grateful to the referee, P. Coppi, for useful comments, to R. Svensson, 
I. V. Igumenshchev, P. B. Ivanov, and especially to A. F. Illarionov for 
discussions. This work was supported by RFFI grant 97-02-16975 and the Swedish 
Natural Science Research Council. I thank M. Abramowicz and the Department of 
Astronomy \& Astrophysics of Chalmers University of Technology, where a part of 
this work was done, for hospitality.

%######################################################################

\appendix
\section{Numerical simulation of gamma-ray transfer}

The radiation field in a plane-parallel atmosphere is axisymmetric
at any height $z$. The photon distribution over energies, $\ep$, and angles 
with 
respect to the $z$-axis, $\theta$, is related to the radiation intensity by
$$
   n(\ep,\theta)=\frac{2\pi \sin\theta I_\ep}{\ep m_{\rm e} c^3}. 
$$
In the numerical simulations, we consider photon energies $0.1<\ep<10$, and 
angles $0<\theta<\pi$. The radiation field at each height is represented
by photon numbers in $N_\ep\times N_\theta=50\times 20$ cells which are
logarithmically spaced in the $\ep$--direction and uniformly spaced in 
the $\theta$--direction. We checked that our results are not significantly 
affected by doubling the grid resolution.

\bigskip
%\begin{center}
\noindent
{\bf (i) $\;\;$ $\gga$ reaction}
%\end{center}
\medskip

\noindent
To simulate the $\gga$ process we need the corresponding effective
cross sections for the interaction between photons from cells
$(i_\ep,i_\theta)$ and $(i_\ep^\prime,i_\theta^\prime)$. Given an
angle between the interacting photons, $\xi$, and 
exact values of their energies, $\ep$ and $\ep^\prime$, one can calculate
the $\gga$ cross section using equations (5) and (7) 
with $\theta$ replaced by $\xi$.
To find the effective cross section for the
interaction between two given cells we use the
Monte-Carlo method. We calculate the cross section for randomly chosen values
$\ep,\theta$ and $\ep^\prime,\theta^\prime$ within the 
cells $(i_\ep,i_\theta)$ and $(i_\ep^\prime,i_\theta^\prime)$ 
and randomly chosen azimuthal angles of the interacting photons. 
We then find the average cross section for $\gga$ interaction between cells
$(i_\ep,i_\theta)$ and $(i_\ep^\prime,i_\theta^\prime)$. Repeating this 
procedure for each pair of cells, we get 
a matrix $\Gamma$ of size $N_\ep\times N_\theta\times N_\ep \times N_\theta$ 
containing the effective cross sections.
The rate of photon depletion in each cell due to $\gga$ reaction is 
given by
\begin{eqnarray}
  \dot{n}_{\gamma\gamma}(i_\ep,i_\theta)=c\; n(i_\ep,i_\theta)
  \sum_{i_\ep^\prime,i_\theta^\prime} 
  \Gamma(i_\ep,i_\theta,i_\ep^\prime,i_\theta^\prime)\; 
  n(i_\ep^\prime,i_\theta^\prime).
\nonumber
\end{eqnarray}

\bigskip
%\begin{center}
\noindent
{\bf (ii) $\;\;$ Compton scattering}
%\end{center}
\medskip

\noindent
We also use the Monte-Carlo method to calculate the effective cross section 
for scattering from cell
$(i_\ep,i_\theta)$ to cell $(i_\ep^\prime,i_\theta^\prime)$. In each 
Monte-Carlo event, we take a random photon energy $\ep$ and angle $\theta$
within the cell $(i_\ep,i_\theta)$, and calculate the total scattering cross
section $\sigma=(1-\beta\cos\theta)\tilde{\sigma}$. 
Here $\beta$ is the plasma velocity (in units of the speed of light) 
directed along the $z$-axis, and ${\tilde\sigma}$ 
is the total cross section in the comoving frame (Jauch \& Rohrlich 1976)
\begin{eqnarray}
  \tilde{\sigma}=
  \frac{3\sigma_{\rm T}}{8{\tilde\ep}}\left[\frac{1}{2}+
  \frac{4}{\tilde{\ep}}-\frac{1}{2(1+2{\tilde\ep})^2}+
  \left(1-\frac{2}{\tilde\ep}-\frac{2}{{\tilde\ep}^2}\right)
  \ln(1+2{\tilde\ep}) \right],
\nonumber
\end{eqnarray}
where ${\tilde\ep}=\gamma(1-\beta\cos\theta)\ep$ is photon energy in the
comoving frame. The photon angle in the comoving frame is given by
\begin{eqnarray}
  \cos{\tilde\theta}=\frac{\cos\theta-\beta}{1-\beta\cos\theta}.
\nonumber
\end{eqnarray}
Then we perform a random scattering 
in the comoving frame. The differential cross section for
Compton scattering is given by (Jauch \& Rohrlich 1976)
\begin{eqnarray}
  \frac{{\rm d}{\tilde\sigma}}{{\rm d}{\tilde\Omega}}=
  \frac{3\sigma_{\rm T}}{16\pi}
  \frac{1}{[1+{\tilde\ep}(1-\mu)]^2}\left[1+\mu^2+\frac{{\tilde\ep}^2(1-\mu)^2}
  {1+{\tilde\ep}(1-\mu)}\right],
\nonumber
\end{eqnarray}
where $\mu={\tilde{\bf\Omega}}\cdot{\tilde{\bf\Omega}}^\prime$, 
${\tilde{\bf\Omega}}$ and ${\tilde{\bf\Omega}}^\prime$ being the photon 
directions
before and after the scattering respectively. The photon energy after the
scattering equals
\begin{eqnarray}
  {\tilde\ep}^\prime=\frac{\tilde\ep}{1+{\tilde\ep}(1-\mu)}.
\nonumber
\end{eqnarray}
Performing the inverse transformation to the lab frame, we get the energy, 
$\ep^\prime$,
and angle, $\theta^\prime$, of the scattered photon, and determine the
corresponding cell $(i_\ep^\prime,i_\theta^\prime)$ into which it has been 
scattered. Repeating $1.5\cdot 10^6$ Monte-Carlo events for each cell 
$(i_\ep.i_\theta)$, we find the distribution of the scattered photons 
throughout all cells $(i_\ep^\prime,i_\theta^\prime)$ and get a matrix 
$C(i_\ep,i_\theta,i_\ep^\prime,i_\theta^\prime)$ containing the required 
effective cross sections. The photon sink and source due to Compton scattering,
${\dot n}_{\rm C}^-$ and ${\dot n}_{\rm C}^+$ respectively, are given by
\begin{eqnarray}
  \dot{n}_{\rm C}^-(i_\ep,i_\theta)=2n_+c\; n(i_\ep,i_\theta)
  {\bar\sigma} (i_\ep,i_\theta),
\nonumber
\end{eqnarray}
\begin{eqnarray}
  \dot{n}_{\rm C}^+(i_\ep,i_\theta)=2n_+c\; 
  \sum_{i_\ep^\prime,i_\theta^\prime} 
  C(i_\ep^\prime,i_\theta^\prime,i_\ep,i_\theta)\; 
  n(i_\ep^\prime,i_\theta^\prime),
\nonumber
\end{eqnarray}
where the bar over $\sigma$ denotes averaging within the cell 
$(i_\ep,i_\theta)$. 
The scattering can affect the total number of photons in the considered energy
range $0.1<\ep<10$ because scattered photons may end up outside this range.
We checked that the code conserves the photon number, i.e., that the decrease 
of the total photon number due to scattering is equal to the photon flux 
through the boundary of the energy interval. 

\bigskip
\bigskip

\noindent
%\begin{center}
{\bf (iii) $\;\;$ Annihilation emission}
%\end{center}
\medskip

\noindent
The probability that an annihilation photon has an angle within 
interval $(\theta,\theta+{\rm d}\theta)$ is given by (see Section 3.2)
\begin{eqnarray}
   {\rm d}P=\frac{\sin\theta {\rm d}\theta}{2\gamma^2(1-\beta\cos\theta)^2}.
\nonumber
\end{eqnarray}
Replacing ${\rm d}\theta$ by $\pi/N_\theta$ and taking 
$\theta=(\pi/N_\theta)(i_\theta-0.5)$, we have the probability $P(i_\theta)$
for an annihilation photon to be in cell $(i_\ep^*,i_\theta)$, where 
$i_\ep^*$ corresponds to the photon energy
$\ep_{\rm ann}=[\gamma(1-\beta\cos\theta)]^{-1}$.
The production rate of annihilation photons is then approximately given by
\begin{eqnarray}
 \dot{n}_{\rm ann}(i_\ep,i_\theta)=2\dot{n}_+^{\rm ann}
                                   P(i_\theta)\delta_{i_\ep i_\ep^*},
\nonumber
\end{eqnarray}
where $\dot{n}_+^{\rm ann}=(3/8)(1-\beta^2)c\sigma_{\rm T}n_+^2$ is the rate 
of pair annihilation.

\bigskip
\bigskip

\noindent
%\begin{center}
{\bf (iv) $\;\;$ The transfer equation }
%\end{center}
\medskip

\noindent
The transfer equation reads 
\begin{eqnarray}
   \frac{\partial n}{\partial t}=-\dot{n}_{\gamma\gamma}-\dot{n}_{\rm C}^-+
   \dot{n}_{\rm C}^++\dot{n}_{\rm ann}-c\cos\theta\frac{\partial n}{\partial z}.
\nonumber
\end{eqnarray}
In the optically thin case (Sections 2.3 and 3.1),
we neglect the terms $\dot{n}_{\rm C}^\pm$ and $\dot{n}_{\rm ann}$,
and we take into account the finite disc dimension by truncating the
radiation at angles $\theta>{\rm arctan}(R/z)$. When simulating 
the optically thick atmosphere in Section 4.1, we account for all terms
in the transfer equation and use the additional equation for $e^\pm$ transfer,
\begin{eqnarray}
  \frac{\partial n_+}{\partial t}=\dot{n}_+^{\gamma\gamma}-\dot{n}_+^{\rm ann}
                               -c\beta\frac{\partial n_+}{\partial z},
\nonumber
\end{eqnarray}
where 
\begin{eqnarray}
  \dot{n}_+^{\gamma\gamma}=\frac{1}{2}\sum_{i_\ep,i_\theta} 
   \dot{n}_{\gamma\gamma}(i_\ep,i_\theta).
\nonumber
\end{eqnarray}
We model the atmosphere as a slab of height $R/2$ above the disc and divide
it into 25 horizontal layers of equal thickness. 
Doubling the number of layers does not strongly affect the results. 
We set the boundary condition at $z=0$ as being a power-law isotropic 
radiation field. The outer boundary condition at $z=R/2$ is free escape of 
the radiation. The main parameter of the problem is the compactness,
$l_1$, defined by equation (2). The other parameter is the spectral index
of the primary radiation, $\alpha$. To find a stationary
state of the atmosphere we take an initially empty 
computational slab $z<R/2$ and follow the evolution
of the radiation field and $e^\pm$ plasma according to the transfer equation
until a stationary state is established.
The time step in the calculations equals  $0.01 R/c$.


\begin{thebibliography}{}

\bibitem[{} {}]{}
Antonucci R. R. J., 1992, in Holt S., Neff S., Urry C.M., eds,  
AIP Conf. Proc. 254, Testing the AGN Paradigm. New York, p. 486

\bibitem[{} {}]{}
Becker P. A., Kafatos M., 1995, ApJ, 453, 83

\bibitem[{} {}]{}
Beloborodov A. M., Illarionov A. F., 1995, ApJ, 450, 64

\bibitem[{} {}]{}
Beloborodov A.M., 1998, ApJ, 496, L105

\bibitem[{} {}]{}
Coppi P.S., Lamb D. Q., 1992, in Paciesas W. S., Fishman G. J., eds, 
AIP Conf. Proc. 265, Gamma-Ray Bursts. New York, p. 257

\bibitem[{} {}]{}
Galeev A. A., Rosner R., Vaiana G. S., 1979, ApJ, 229, 318

\bibitem[{} {}]{}
Ghisellini G., Celotti A., George I. M., Fabian A. C., 1992, MNRAS, 258, 776

\bibitem[{} {}]{}
Gould R. J., Schr\'{e}der G. P., 1967, Phys. Rev., 155, 1404 

\bibitem[{} {}]{}
Guilbert P. W., Fabian A. C., Rees M. J., 1983, MNRAS, 205, 593 

\bibitem[{} {}]{}
Guilbert P. W., Stepney S., 1985, MNRAS, 212, 523

\bibitem[{} {}]{}
Gurevich L. E., Rumyantsev A. A., 1965, Sov. Physics -- JETP, 20, 1233

\bibitem[{} {}]{}
Herterich K., 1974, Nat, 250, 311

\bibitem[{} {}]{}
Icke V., 1989, A\&A, 216, 294

\bibitem[{} {}]{}
Illarionov A. F., Krolik J. H., 1996, ApJ, 469, 698

\bibitem[{} {}]{}
Jauch J. M., Rohrlich F., 1976, The Theory of Photons and Electrons.
 Springer, New York

\bibitem[{} {}]{}
Katz J.I., 1982, ApJ, 260, 371

\bibitem[{} {}]{}
Li H., Liang E.P., 1996, ApJ, 458, 514

\bibitem[{} {}]{}
McNaron-Brown K. et al., 1995, ApJ, 451, 575 

%\bibitem[{} {}]{}
%O'Dell S. L., 1981, ApJ, 243, L147

%\bibitem[{} {}]{}
%Phinney S., 1982, MNRAS, 198, 1109

\bibitem[{} {}]{}
Phinney S., 1983, PhD thesis, University of Cambridge

\bibitem[{} {}]{}
Phinney S., 1987, in Zenus J. A., Pearson T. J., eds, Superluminal Radio 
Sources. Cambridge Univ. Press, p. 301

\bibitem[{} {}]{}
Reeves J. N., Turner M. J. L., Ohashi T., Kli T., 1997, 292, 468

\bibitem[{} {}]{}
Rybicki G.B., Lightman A.P., 1979, Radiative Processes in Astrophysics,
Wiley, New York

\bibitem[{} {}]{}
Sikora M., Wilson D. B., 1981, MNRAS, 197, 529

%\bibitem[{} {}]{}
%Shakura N. I., Sunyaev R. A., 1973, A\&A, 24, 337

\bibitem[{} {}]{}
Shapiro S. L., Lightman A. P., Eardley D. M., 1976, ApJ, 204, 187

\bibitem[{} {}]{}
Svensson R., 1987, MNRAS, 227, 403 

\bibitem[{} {}]{}
Svensson R., 1996, A\&AS, 120, 475  

%\bibitem[{} {}]{}
%Tsuruta S., Kellen M., 1995, ApJ, 453, L9 

\bibitem[{} {}]{}
Zdziarski A. A., 1984, A\&A, 134, 301 


\end{thebibliography}
\end{document}